\documentclass[12pt,preprint]{aastex}

\shorttitle{A Comparative Study of Complex Eruptions}

\shortauthors{Liu et al.}

\begin{document}

\title{A Comparative Study of 2017 July and 2012 July Complex Eruptions: Are Solar Superstorms ``Perfect Storms" in Nature?}

\author{Ying D. Liu\altaffilmark{1,2}, Xiaowei Zhao\altaffilmark{1,2}, Huidong Hu\altaffilmark{1}, 
Angelos Vourlidas\altaffilmark{3}, and Bei Zhu\altaffilmark{1,2}} 

\altaffiltext{1}{State Key Laboratory of Space Weather, National Space 
Science Center, Chinese Academy of Sciences, Beijing, China; liuxying@swl.ac.cn}

\altaffiltext{2}{University of Chinese Academy of Sciences, Beijing, China}

\altaffiltext{3}{The Johns Hopkins University Applied Physics Laboratory, Laurel, MD 20732, USA} 

\begin{abstract}

It is paramount from both scientific and societal perspectives to understand the generation of extreme space weather. We discuss the formation of solar superstorms based on a comparative study of the 2012 July 23 and 2017 July 23 eruptions. The first one is Carrington-class, and the second could rival the 1989 March event that caused the most intense geomagnetic storm of the space age. Observations of these events in the historically weak solar cycle 24 indicate that a solar superstorm can occur in any solar cycle and at any phase of the cycle. Recurrent patterns are identified in both cases, including the long-lived eruptive nature of the active region, a complex event composed of successive eruptions from the same active region, and in-transit interaction between the successive eruptions resulting in exceptionally strong ejecta magnetic fields at 1 AU. Each case also shows unique characteristics. Preconditioning of the upstream solar wind leading to unusually high solar wind speeds at 1 AU is observed in the first case whereas not in the latter. This may suggest that the concept of ``preconditioning" appears to be necessary for making a Carrington-class storm. We find a considerable deflection by nearby coronal holes in the second case but not in the first. On the basis of these results, we propose a hypothesis for further investigation that superstorms are ``perfect storms" in nature, i.e., a combination of circumstances that results in an event of unusual magnitude. Historical records of some extreme events seem to support our hypothesis.

\end{abstract}


\section{Introduction}

Coronal mass ejections (CMEs) are large-scale expulsions of mass and magnetic field from the solar corona and have been recognized as primary drivers of space weather. An extreme case of CMEs is termed a low-probability, high-consequence event, otherwise called a solar superstorm. It poses severe risks to critical infrastructures of the modern society if it hits the Earth \citep{report08, cannon13, oughton17, riley18}. Studies of solar superstorms are of fundamental importance from both scientific and societal perspectives. 

There are many parameters that could be used to assess the extremeness or severity of a solar storm, such as flare intensity, CME speed and fluxes of solar energetic particles \citep{cliver04, riley12, schrijver12}. An event may be extreme with respect to some parameters, but only moderate relative to others. Here we use the solar wind transient speed ($v$) and magnetic field (essentially the southward component $B_{\rm s}$) at 1 AU, with a focus on the potential of CMEs to create large geomagnetic storms. This is because $vB_{\rm s}$, the solar wind dawn-dusk electric field, controls the rate of the solar wind energy coupling to the terrestrial magnetosphere \citep{dungey61}, and $nv^2$ (where $n$ is the solar wind density), the solar wind momentum flux, governs the compression extent of the magnetosphere. Clearly, to increase geoeffectiveness requires simultaneous enhancement of the southward magnetic field, the solar wind speed and the plasma density, in order of importance. The question thus boils down to how these three parameters, especially the first two, can be extremely enhanced simultaneously. 

One may first think of the most extreme eruption that could be produced by the largest, most energetic active region possible, given the current age of the Sun \citep[e.g.,][]{schrijver12, shibata13}. A statistical analysis indicates that the maximum speed of CMEs near the Sun cannot be much higher than 3000 km s$^{-1}$ \citep{yashiro04}. \citet{gopal05} argue that this speed limit of $\sim$3000 km s$^{-1}$ implies an upper bound to the maximum energy available to a CME from active regions. It would be interesting to perform numerical experiments to see what would be the largest possible magnetic field and speed at 1 AU that could be generated by the most powerful eruption at the Sun, given this limitation on the maximum free energy of active regions and a typical ambient solar wind background through which the eruption propagates. In situ measurements indicate that most CMEs end up with speeds and magnetic fields comparable to those of the ambient solar wind when they reach 1 AU \citep[e.g.,][]{liu05, jian06, richardson10}. Therefore, the evolution of CMEs in interplanetary space is also key to determining their severity at the Earth, in addition to the initial active region conditions on the Sun. Interactions with the ambient solar wind will decelerate fast CMEs, and an initially faster CME would have a stronger deceleration \citep{gopal00, reiner07, liu13, liu16}. The ejecta internal magnetic field (and density as well) will also drastically decrease because of expansion of the ejecta in interplanetary space \citep{kumar96, liu05, liu06}. We speculate that a single isolated extreme eruption supported by the current state of the Sun might not be able to result in the extreme magnetic field and speed that we can imagine at 1 AU.  

Based on the observations of the 2012 July 23 complex CME, \citet[][hereinafter referred to as Paper I]{liu14} suggest a ``perfect storm" scenario for the generation of an extreme event at 1 AU: a preconditioning effect to prevent deceleration (i.e., one or several earlier eruptions clear the path), plus in-transit interaction between later closely launched eruptions to preserve the magnetic field (as well as the density). With this ``perfect storm" scenario all three parameters can be significantly enhanced at the same time. In the 2012 July 23 case, the magnetic field inside the ejecta and the solar wind transient speed both reach record values at 1 AU, i.e., 109 nT and 2246 km s$^{-1}$ respectively. No reliable plasma density measurements exist for this event, but there are indications that the density may have been 100 cm$^{-3}$ or higher at 1 AU (see Paper I). This new view on the generation of an extreme event emphasizes the crucial importance of CME interplanetary evolution for space weather, here CME-CME interactions as previously postulated \citep[e.g.,][]{gopalswamy01, burlaga02, farrugia04, lugaz05, liu12}. The preconditioning effect could be considered as indirect interaction between CMEs of interest. The ``perfect storm" mechanism may be one of the best (if not the only) ways to produce an ejecta magnetic field larger than 100 nT and a solar wind transient speed higher than 2000 km s$^{-1}$ at 1 AU. 

In addition to the ejecta, the sheath region between the ejecta and its driven shock can also carry enhanced southward magnetic fields produced by shock compression of preexisting quiet solar wind fields \citep[][and references therein]{kilpua17}. \citet{tsurutani92} suggest that the sheath fields are at least as geoeffective as the ejecta fields. Although large sheath fields have been observed, they are usually very brief and turbulent \citep[e.g.,][]{duston77, skoug04}. Therefore, sheath fields per se may not be as effective as the ``perfect storm" mechanism in generating extreme events. Reinforcing factors may be needed, e.g., a preceding ejecta providing source southward fields for shock compression, or a preconditioning effect to enhance the shock speed. However, these essentially fall into the ``perfect storm" scenario (see more discussions in Section 4). Reexamination of the data in the literature on the cause of large geomagnetic storms indeed indicates some reinforcing factors. For example, the Figure~1 of \citet{tsurutani92} shows a very low density ahead of the shock and a high-speed stream compressing the ejecta from behind.

On 2017 July 23 near the end of the declining phase of the historically weak solar cycle 24, exactly five years after the 2012 extreme event, the Sun produced another powerful CME that gave rise to an ejecta with exceptionally strong internal magnetic fields at 1 AU. In this paper we perform a comparative study of these two solar superstorms\footnote{Throughout the paper the term ``solar superstorms" refers to extreme solar wind disturbance at 1 AU, and should not be confused with ``geomagnetic superstorms". We infer that the two events would have caused geomagnetic superstorms if they had been Earth directed.}, in an attempt to address a critical question concerning the formation of extreme events: are solar superstorms ``perfect storms" in nature? The paper is organized as follows. We summarize some of the results on the 2012 July event from Paper I relevant to the present work in Section 2, and then describe in detail our analysis of multi-point imaging and in situ observations of the 2017 July event in Section 3. The results from the two cases are compared and discussed in Section 4. This study provides a benchmark theory or hypothesis for the generation of extreme space weather events that can be further investigated and assessed.   

\section{The 2012 July 23 Event} 

The 2012 July 23 extreme CME has attracted significant attention because of its unusually high solar wind speed and extremely strong ejecta magnetic field observed near 1 AU \citep[e.g.,][]{liu14, liu17a, russell13, baker13, temmer15, cash15, riley16, zhang16, zhu16}. In Paper I we have analyzed multi-point imaging observations in connection with in situ measurements with a focus on the formation of the extreme storm. Readers are directed to Paper I for a detailed discussion of the analysis. Here we summarize some of the results from Paper I relevant to the present work and make further clarifications wherever necessary for our discussions. 

Figure~1 shows the positions of three widely separated spacecraft that have imaging capabilities, including the Solar and Heliospheric Observatory \citep[SOHO;][]{domingo95} upstream of the Earth at L1 and the two Solar Terrestrial Relations Observatory \citep[STEREO;][]{kaiser08} spacecraft (labeled as ``A" for STEREO Ahead and ``B" for STEREO Behind). The event originated from AR 11520, which was not particularly large but kept erupting for a long time. The eruptions from this active region include the July 12 event, which was Earth directed and produced a relatively intense geomagnetic storm \citep[e.g.,][]{hess14, hu16}, and the July 19 one, which helped precondition the upstream solar wind for the July 23 complex CME. During the time of the 2012 July 23 eruptions, STEREO A and B were 0.96 AU and 1.02 AU from the Sun and 121.2$^{\circ}$ west and 114.8$^{\circ}$ east of the Earth, respectively. The source region (S15$^{\circ}$W133$^{\circ}$) was about 12$^{\circ}$ west of the central meridian in the view of STEREO A. The event started around 02:20 UT with two consecutive prominence eruptions separated by about 10 - 15 minutes, as can be seen in the observations of STEREO B. Soon the two eruptions merged near the Sun, and no definite twin-CME signatures could be identified in the coronagraph observations from any of the three spacecraft. At the same time a shock signature developed around the complex event, which can be well fitted by a spherical structure in all the three views \citep{liu17a}. The spherical modeling of the shock gives a propagation direction of about 8$^{\circ}$ west of STEREO A, which does not deviate much from the solar source longitude (see Figure~1 left). The peak speed of the CME nose is about 3000 km s$^{-1}$.      

A forward shock passed STEREO A at 20:55 UT on July 23, behind which two interplanetary CMEs (ICMEs) could be identified. The transit time to STEREO A is only about 18.6 hours, similar to that (17.6 hours) of the 1859 Carrington event \citep{carrington1859}, a well-known example of extreme space weather before the space era. Readers are directed to \citet{cliver04} and \citet{cliver90} for the transit time and storm level rankings of some earlier events. The peak speed behind the shock is 2246 km s$^{-1}$, and the maximum magnetic field is 109 nT in the early part of the complex ejecta. Both values are among the few largest ever measured near 1 AU. Electron measurements suggest that the plasma density in the first ICME and the interaction interface between the two ICMEs may have reached 100 cm$^{-3}$ or higher. The average transit speed of the shock (about 2150 km s$^{-1}$) is close to the peak solar wind speed at STEREO A, so the event did not slow down much during the transit. The magnetic field has a sustained southward component with a maximum value of 52 nT inside the second ICME. If the event had hit the Earth, it would have likely generated a geomagnetic storm comparable to that of the Carrington event.

Upstream of the shock, the solar wind density is as low as 1 cm$^{-3}$ and the magnetic field predominantly lies along the radial direction from the Sun. This is the so-called preconditioning effect caused by the earlier eruptions including the July 19 one. The radial magnetic field would lessen the field line tension force to be experienced by the later complex event, and the tenuous upstream plasma will significantly reduce the solar wind drag \citep[for the definition of solar wind drag see][]{vrsnak13}. The radial orientation of the field also helps drain the solar wind plasma and thus maintain the low-density state for a longer time. The observed minimal slowdown of the event can be explained by this preconditioning effect. The in-transit interaction between the two closely launched eruptions could inhibit the expansion of the ejecta, helping preserve the large magnetic field and density inside the ejecta. It is likely that the passage of the shock driven by the second CME through the first one would also enhance the preexisting plasma density and magnetic field. This ``perfect storm" scenario can give rise to an event with unusually high solar wind speed and extremely enhanced magnetic field and density near 1 AU, besides the initial active region conditions on the Sun. 

\section{The 2017 July 23 Event} 

Another impressive event occurred on the fifth anniversary of the 2012 extreme CME near the end of the declining phase of solar cycle 24. The general locations of the two STEREO spacecraft are now switched (see Figure~1 right). They were 132$^{\circ}$ east and 127$^{\circ}$ west of the Earth and 0.97 AU and 1.02 AU from the Sun, respectively. STEREO B lost contact with the Earth in 2014 October, so no data are available from it. The MAVEN spacecraft around Mars, which was 178$^{\circ}$ east of the Earth and 1.64 AU from the Sun, provides another vantage point of in situ measurements. The source region, AR 12665, was roughly S06$^{\circ}$E88$^{\circ}$ relative to STEREO A when the event occurred. Therefore, it was an east-limb event for STEREO A. As will be shown below, this is also a complex event composed of two consecutive eruptions. Again, the active region was long-lived, not particularly large but kept erupting. Among the eruptions from this active region, the July 14 event produced a geomagnetic storm only with its flank encountering the Earth, and another one occurred just a few hours before the July 23 event (CME0 in Figure~1). Both the July 23 complex CME and CME0 are deflected from the active region.     

\subsection{Multi-Point Imaging Observations}

Figure~2 displays the solar source and evolution of the complex CME observed from STEREO A. The EUV images present clear evidence of two successive eruptions. The first one occurred around 03:26 UT, as can be seen from the flaring in 304 \AA\ (Figure~2d) and a subsequent coronal wave in 195 \AA\ (Figure~2a). About 10 minutes later, a prominence erupted (Figure~2e), and another coronal wave, which is much stronger than the first wave, emerged from the same place (Figure~2b). This is the second eruption. A jet-like eruption (Figure~2c) and an extended flaring phase (Figure~2f) are observed, which indicates the energetic nature of the active region. The orientations of the erupting prominence and the post-eruption arcade are largely vertical, so the CME flux rope may have a big tilt angle. Later we see the flaring spread westward (Figure~2f), which may imply a change in the flux-rope orientation. CME0 was associated with another prominence eruption from the same active region a few hours earlier, which exhibited a non-radial motion (not shown here). Twin-CME signatures are seen in COR1 around 04:20 UT (Figure~2g), with the first one (CME1) much weaker than the second one (CME2). There is a structure in COR2 at 05:09 UT (Figure~2h), which seems separated from the main body of CME2, but it is not clear if this is part of CME1. Soon a complex CME was formed without clear twin-CME signatures (Figure~2i). A global shock wave, a faint edge ahead of the CME front, developed at about the same time.       

Figure~3 shows multi-point coronagraph observations and modeling of the CME and shock. The event appears as a full halo in both SOHO and STEREO A, which indicates that a deflection may have occurred, so it would hit both STEREO A and Mars (see Figure~1 right). We employ a graduated cylindrical shell (GCS) method to model the CMEs, which assumes a rope-like morphology with two ends anchored at the Sun \citep{thernisien06}. The twin CMEs cannot be fit separately, because they occurred so close in time and space and only parts of them were observed. We apply the model to the complex CME as if it were a single eruption. It is very difficult to fit the CME precisely, and the observed CME is always much wider than the model for SOHO (see Figure~3e). Note that the COR2 images from STEREO A provide a strong constraint on the fit. The difficulty for an accurate fit gives another indication of consecutive eruptions within the complex event. 

The shock, however, can be fitted fairly well by a spherical model in both views at later times, similar to what we have seen in the 2012 July case \citep{liu17a}. The propagation direction of the shock from the modeling is about 170$^{\circ}$ east of the Earth and 25$^{\circ}$ south, which is between the longitudes of Mars and STEREO A. The complex event may have been deflected by about 50$^{\circ}$ from the solar source longitude (see Figure~1 right). The propagation direction does not change much in the time series of the shock modeling, so the deflection may mainly take place in the low corona. The GCS fit of CME0 (not shown here) gives a propagation direction of about 155$^{\circ}$ west of the Earth and 13$^{\circ}$ north, so CME0 is less deflected (only about 15$^{\circ}$). The difference between the propagation directions of the complex CME and CME0 and the narrow width of CME0 suggest that they may not have interacted much (if there is any interaction). A linear fit of the resulting distances gives a speed of about 2300 km s$^{-1}$ for the complex shock and about 870 km s$^{-1}$ for CME0 near the Sun.                   

To investigate why the CMEs deviated from the source longitude, we examine the EUV observations and the coronal magnetic field configuration around the active region during its rotation across the visible solar disk of STEREO A (see Figure~4). The active region persisted for a long time. Here we present an EUV image from July 30. The coronal magnetic fields are obtained with a potential-field source-surface model \citep[e.g.,][]{schatten69, altschuler77}, a global extrapolation that is assumed not to change much during a solar rotation period. Coronal holes with open field lines, which can last several solar rotations, are observed on the immediate east of the active region. The nearby open field lines lean westward above the active region. The coronal holes and these particularly oriented open fields may deflect the CMEs in the corona towards the directions of Mars and STEREO A. Deflection of CMEs by coronal holes has been shown in previous studies \citep[e.g.,][]{gopal09}. CME0 is narrow, which might help explain its less deflection than the complex CME. It should be stressed, again, that the deflections must have occurred in the lower corona because the propagation directions of the complex shock and CME0 are roughly constant in the outer corona. 

\subsection{Multi-Point In Situ Measurements}

The in situ measurements at STEREO A are shown in Figure~5. A pair of shocks, which are step-like and only 9 minutes apart, are observed around 17:54 UT on July 24 with a speed of about 660 km s$^{-1}$ for the first one and about 840 km s$^{-1}$ for the second one. Two ICMEs can be identified from the depressed proton temperature and smooth magnetic field, which presumably correspond to the two successive eruptions near the Sun\footnote{We examine coronagraph images from SOHO and STEREO A for a few days before the complex eruptions to check if there could be any other candidate for ICME1, but do not find one with clear evidence that can arrive at either of STEREO A and Mars. CME0 can be excluded too, given its propagation direction and narrow width (about 20 - 30$^{\circ}$).}. The ICME intervals are consistent with reductions in the proton $\beta$, which is well below 0.01 inside ICME1 and generally below 0.2 inside ICME2 (not shown here). As can be seen from the speeds, ICME2 was still compressing ICME1 near 1 AU. The magnetic field strength inside ICME1 is as high as 68.7 nT and the peak southward field component is 60.7 nT, both of which are unusually large values at 1 AU. The $B_{\rm N}$ component is purely negative (southward) inside ICME1, which appears to be consistent with the orientation of the post-eruption arcade. The magnetic field within ICME2 is primarily along the radial direction from the Sun, which may have been caused by the fast expansion of the ejecta. The expansion speed of ICME2 relative to its center is about 200 km s$^{-1}$, much larger than the Alfv\'en speed inside ICME2 which is generally below 100 km s$^{-1}$. The plasma flow dominates over the magnetic field inside ICME2, so the field line is stretched to be along the radial direction.         

Given the enormous southward magnetic field and its persistence for about 5.1 hours, the event would produce an immense geomagnetic storm if it hit the Earth. We evaluate the $D_{\rm st}$ index, a measure of geomagnetic storm intensity, using two empirical formulas based on the solar wind measurements \citep{burton75, om00}. The difference between the two resulting $D_{\rm st}$ profiles is owing to different assumptions on the decay of the terrestrial ring current in the two formulas. The minimum $D_{\rm st}$ value is about $-590$ and $-330$ nT, respectively. The upper limit ($-590$ nT) rivals the most severe geomagnetic storm of the space era caused by the 1989 March event \citep[e.g.,][]{cliver04}, which left six million people without electricity for up to 9 hours in Canada. Even the lower limit ($-330$ nT) would place the event among the top 20 superstorms since 1932 \citep{cliver04, gonzalez11}. The observed solar wind speed is much lower than that of the 2012 July event. If the solar wind speed were comparable, the $D_{\rm st}$ index would significantly increase.   

Figure~6 gives the in situ measurements at Mars (1.64 AU). The density and speed profiles are generally similar to those at STEREO A (in particular the double peaks in the speed), so the complex event may have arrived at both STEREO A and Mars. This result corresponds with what the coronagraph imaging observations imply. We cannot interpret the data in a quantitative way, because the temporal resolution is low and the data may be contaminated by the Martian plasma. A forward shock probably arrived at Mars around 19:10 UT on July 25, as can be identified from the peak in the high-energy particles (known as energetic storm particles trapped around the shock). If we propagate the shock outward with the speed of 840 km s$^{-1}$ observed at STEREO A, the shock would arrive at Mars around 03:00 UT on July 26. This is roughly consistent with the observed shock arrival time at Mars, but note that the arrival at Mars should be earlier as the shock nose is more towards Mars (see Figure~1 right). There is a decrease in the electron flux on July 28, which may indicate the location of the leading edge of ICME2. The high-energy particles may be excluded from ICME2 because of the magnetic field inside the ejecta. This interpretation seems to agree with the reduced solar wind density near the same time.           

\subsection{Interplanetary Propagation}

Figure~7 shows the radio dynamic spectrum from STEREO A. Three consecutive type III radio bursts are observed on July 23, which agrees with the multiple successive eruptions (including the jet-like one) at the Sun. Type II bursts, which are well-known shock signatures, are seen as a few faint but discernible patches at fundamental and harmonic frequencies. We use an analytical model proposed by \citet{liu17b} to characterize the shock propagation and simulate the frequency drift of the type II burst. The complex shock is assumed to start with an initial speed near the Sun, move with a constant deceleration ($a$) for a time period ($t_{\rm a}$) before reaching STEREO A, and thereafter travel with a constant speed. The model only requires three parameters: the initial speed that can be obtained from shock modeling near the Sun (2300 km s$^{-1}$), and the shock speed (840 km s$^{-1}$) and arrival time (17:54 UT on July 24) that are known from in situ measurements at STEREO A. The Sun-to-1 AU propagation profile can then be uniquely set. The distances from the analytical model are converted to frequencies using the density model of \citet{leblanc98} with a nominal 1-AU density of 10 cm$^{-3}$. The resulting frequency-time curve reproduces the observed frequency drift and successfully connects two well-separated type II bands. 

The analytical model yields the deceleration $a=-32.1$ m s$^{-2}$, the time for deceleration $t_{\rm a}=12.6$ hours, and the cessation distance of deceleration $r_{\rm a}=0.48$ AU. These parameters are not very different from those of a typical fast CME/shock \citep{liu08, liu13}, so the complex event may not be associated with a considerable preconditioning effect that we have seen in the 2012 July case. This explains the significant slowdown of the shock (2300 km s$^{-1}$ near the Sun versus 840 km s$^{-1}$ at 1 AU). The shock arrival time at Mars suggests that the shock speed may not have changed much during the transit from 1 AU to Mars. We also look at data from Wind, but find no clear type II burst in the radio dynamic spectrum and no shock arrival in the solar wind measurements near the Earth.      

\section{Comparison and Discussion}

Two solar superstorms occurred in the historically weak solar cycle 24, one that is Carrington-class, and the other that may rival the 1989 March event. We have investigated the formation of the 2012 July 23 case in Paper I and of the 2017 July 23 case in the present work. Different from historical extreme events, they were covered by modern remote-sensing and in situ observations from multiple well-aligned spacecraft. As mentioned in Paper I, only with multipoint remote-sensing and in situ observations can we obtain a complete view of what was happening around the Sun and in the heliosphere, and discover how a combination of coronal and interplanetary conditions produced an extreme event at 1 AU. Below we compare these two solar superstorms and discuss the nature of extreme events.

Similarities are found in both cases. First, both of the source active regions are not particularly large but are long-lived and keep erupting. This is distinct from our impression that an extreme event should be produced by an extreme active region. Second, both are complex events composed of two closely launched eruptions from the same active region. In each case, the two consecutive eruptions merge quickly and individual signatures cannot be identified in the coronagraph images. Third, in both cases the in-transit interaction between the two successive eruptions inhibits the expansion of the complex ejecta, which gives rise to immense magnetic fields (including the southward component) inside the ejecta at 1 AU. Both would have produced geomagnetic superstorms if they had hit the Earth. 

Differences are also observed between the two cases. A primary difference is that a significant preconditioning of the upstream solar wind is seen in the 2012 July case while not in the 2017 July event. The latter exhibits interplanetary propagation characteristics more of a typical fast CME. Without a noteworthy preconditioning effect the 2017 July event shows a substantial deceleration, i.e., about 2300 km s$^{-1}$ near the Sun versus about 800 km s$^{-1}$ at 1 AU. Therefore, the concept of ``preconditioning" proposed in Paper I seems key to making a Carrington-class storm. It also stresses the critical importance of CME interplanetary evolution, i.e., how the extreme properties of a CME can be retained when far away from the Sun. Another difference is that there may have been a large deflection by nearby coronal holes in the second case but not in the first. The latter is an east-limb event for STEREO A but deflected towards Mars and STEREO A. These differences suggest that, while common recurrent patterns are present in both cases, each event does have its own unique characteristics.   

Based on this comparative study, we propose a hypothesis that a large fraction of solar superstorms, if not all of them, are ``perfect storms" in nature. With the limited set of events we identify recurrent patterns, including the long-lived eruptive nature of the source active region, a complex event composed of successive eruptions from the same active region, and in-transit interaction between the successive eruptions to inhibit the ejecta expansion. Although a substantial preconditioning effect is not observed in one of the cases, we envision that it may also be recurrent for active regions erupting repeatedly as will be discussed below. All these conform to the ``perfect storm" idea presented in Paper I, i.e., a combination of circumstances results in an event of unusual magnitude. This hypothesis is similar to the suggestion that solar superstorms, at least some aspects of them, may be homologous in nature.

Observations of these two events in a very weak solar cycle indicate that a solar superstorm can occur in any solar cycle and at any phase of the cycle. Previous solar cycles give similar indications \citep[e.g.,][]{kilpua15}, i.e., extreme events could occur in weak cycles and the Sun near a cycle minimum can launch superstorms. Observations of the two events also suggest that ``perfect storms" are not as rare as the phrase implies. The ``perfect storm" scenario is further extended by \citet{liu15} to include any combination of circumstances that can aggravate the situation drastically, such as pileup of events, pre-event plasma rarefactions and magnetic field line stretching, shock enhancement of preexisting ejecta magnetic fields, and fast streams causing compressions from behind. Complex events with such combinations are frequent enough in the solar wind to increase the possibilities of ``perfect storm" scenarios. The complex events discussed here somewhat resemble compound streams (defined as streams resulting from the interaction of two or more fast flows), which can give rise to large geomagnetic storms \citep{burlaga87}. Note that the ``perfect storm" hypothesis does not require all solar superstorms to hold the same patterns or combinations. Each solar superstorm can have its own unique features, as we have seen in the present two cases. As long as they are formed from a combination of circumstances, they can be called ``perfect storms".

The statement that superstorms are ``perfect storms" may sound overstated at a preliminary thought. First, our definition of a superstorm focuses on the solar wind transient speed and ejecta magnetic field at 1 AU, not flare intensity, energetic particle flux, etc, although there may be correlations with them. Second, we do not exclude completely the possibility that an isolated extreme eruption from an extreme active region propagating in a typical solar wind background ends up with an extreme storm at 1 AU. The Sun at its current age might have this capability although we do not see clear evidence. Third, the purpose of this paper is mainly to set up a hypothesis for further test and investigation. However, large active regions often last a long time and erupt repeatedly, so there are good reasons to take this hypothesis seriously. For such an active region the scenario that we have found in the 2012 July case could come up again without much difficulty: the earlier eruptions from the active region can precondition the preexisting ambient solar wind, and the later successive ones from the same active region, while moving through the preconditioned wind, may interact en route to 1 AU. This situation could also happen for active regions that are eruptive and close to each other. 

Historical records of some extreme events seem to support our hypothesis. Around the 1859 Carrington event, which also occurred in a weak solar cycle, long-duration brilliant aurorae were observed from August 28 through September 3. \citet{green06} suggest that the active region erupted successively and two closely spaced CMEs interacted in interplanetary space leading to the extreme nature of the Carrington event. Similarly, long-duration low-latitude auroral displays were seen for almost 9 nights during 1770 September. \citet{hayakawa17} indicate that multiple consecutive eruptions possibly from the same active region, which may have resulted in the ``perfect storm" situation as we propose in Paper I, must have occurred for such a long duration of auroral observations. A more recent case, the ultrafast CME in 1972 August, reached the Earth in only 14.6 hours, the shortest transit time ever observed. \citet{knipp18} suggest that two earlier eruptions from the same active region likely cleared the path for the subsequent ultrafast event, exactly our ``preconditioning" idea discussed in Paper I and here in the current work. We speculate that some of the 2003 Halloween storms may also be ``perfect storms", including the preconditioning effect and/or CME-CME interactions, because several large active regions continuously erupted for a long time \citep[e.g.,][]{gopal05, skoug04}.   

We think that the hypothesis is valuable in that it points out the reality of extreme events from today's Sun. They generally are not simple and can involve many reinforcing factors. While the paper appears to stress the importance of interplanetary evolution of CMEs, the ``perfect storm" concept also includes solar aspects, such as the persistent eruptive nature of the active region and the simultaneous presence of several big active regions on the disk. A particular instance is that widespread magnetic connectivity between concurrent active regions leads to sympathetic eruptions \citep[e.g.,][]{schrijver11, wang16, jin16}. A nearly ``global storm" could be made under this solar-side ``perfect storm" scenario. This may have occurred during the 2003 Halloween period, which had a few large active regions including one that emerged quickly and became connected to the preexisting eruptive complex ones (J. G. Luhmann 2018, private communication). Perhaps it is the combination of the solar and interplanetary aspects of the ``perfect storm" concept that can be the worst.

\acknowledgments The research was supported by NSFC under grant 41774179 and the Specialized Research Fund for State Key Laboratories of China. A.V. was supported by NASA LWS TR\&T program NNX15AT42G and grant NNX16AH70G. We acknowledge use of the excellent data sets from STEREO, SOHO, MAVEN and Wind, and are grateful to the experimenters and providers who made the data publicly available. We thank Janet G. Luhmann for discussions on the work and Christina O. Lee for sending us the MAVEN data.

\clearpage

\begin{figure}
\centerline{\includegraphics[width=20pc]{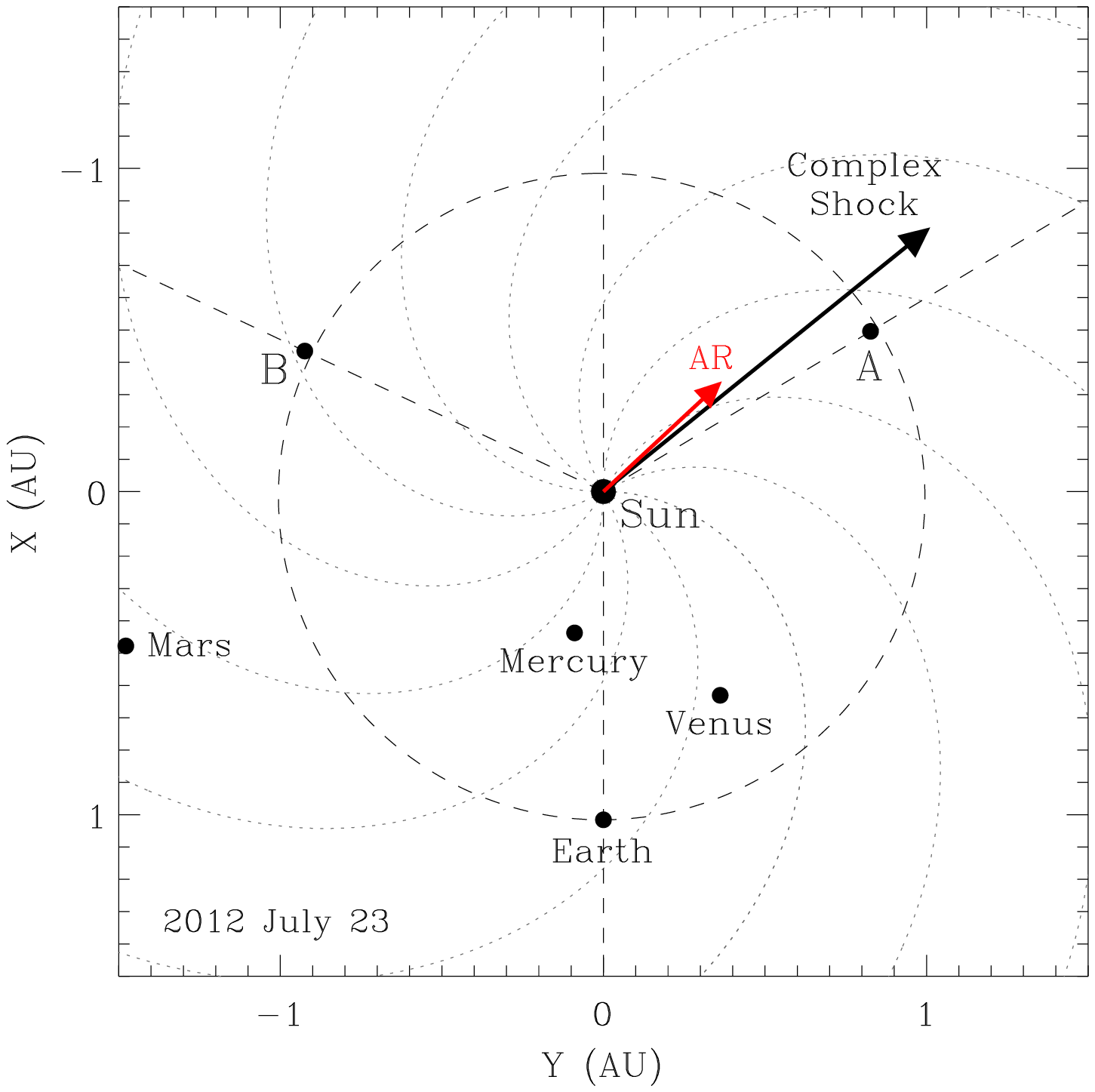}\hspace{1pc}\includegraphics[width=20pc]{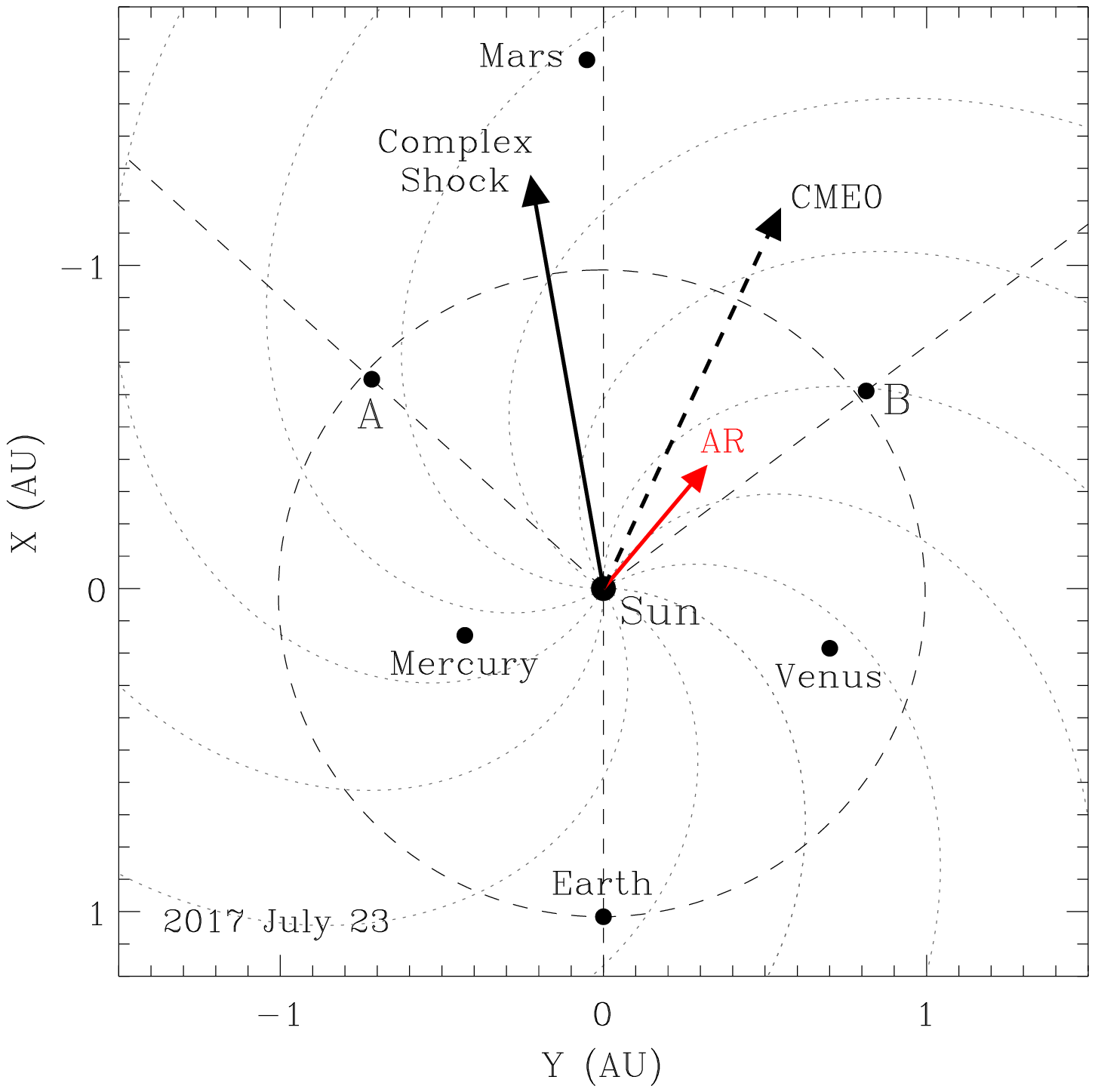}}
\caption{Positions of the spacecraft and planets in the ecliptic plane on 2012 July 23 (left) and 2017 July 23 (right). The black solid arrow marks the propagation direction of the complex shock, and the red solid arrow shows the longitude of the associated active region (AR). Also indicated is the direction of an earlier eruption (CME0) on 2017 July 23. The dashed circle represents the orbit of the Earth, and the dotted lines denote the spiral interplanetary magnetic fields.}
\end{figure}

\clearpage

\begin{figure}
\epsscale{0.75} \plotone{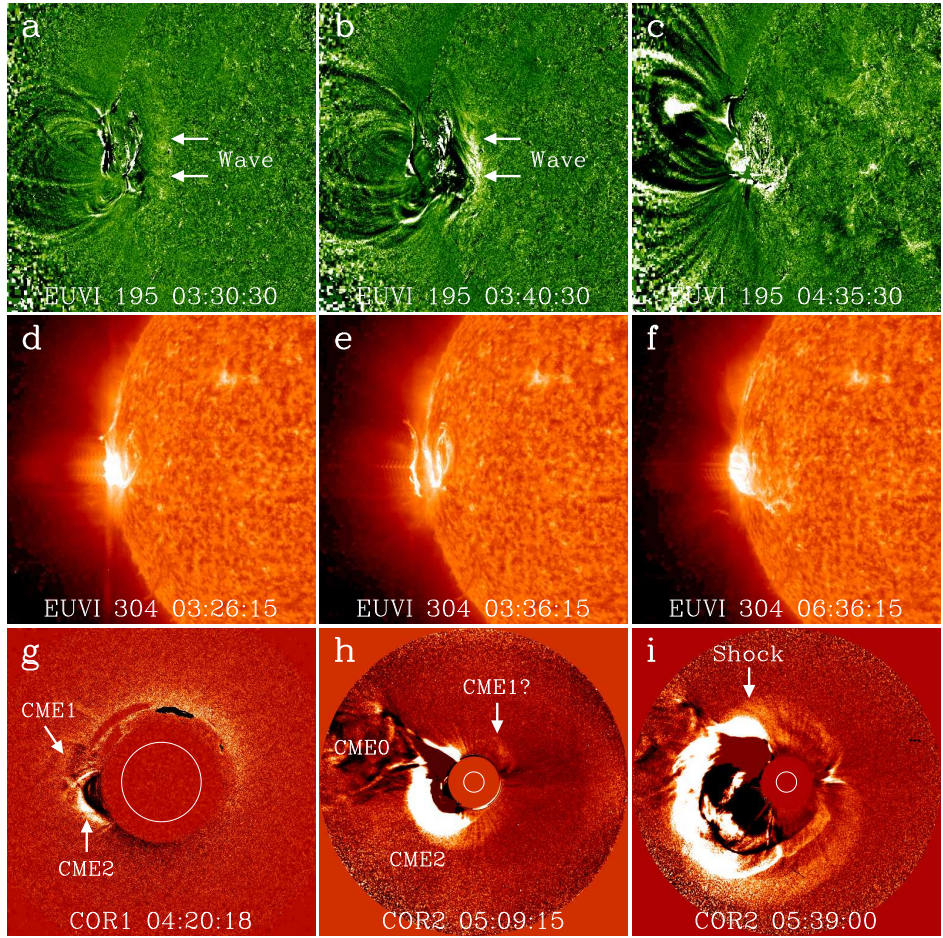} 
\caption{Solar source and evolution of the 2017 July 23 event viewed from STEREO A. (a-c) Running-difference images from EUVI at 195 \AA. (d-f) EUVI images at 304 \AA. (g-i) Running-difference coronagraph images from COR1 and COR2. A final complex CME and shock is formed from the merging of CME1 and CME2.}
\end{figure}

\clearpage

\begin{figure}
\epsscale{0.9} \plotone{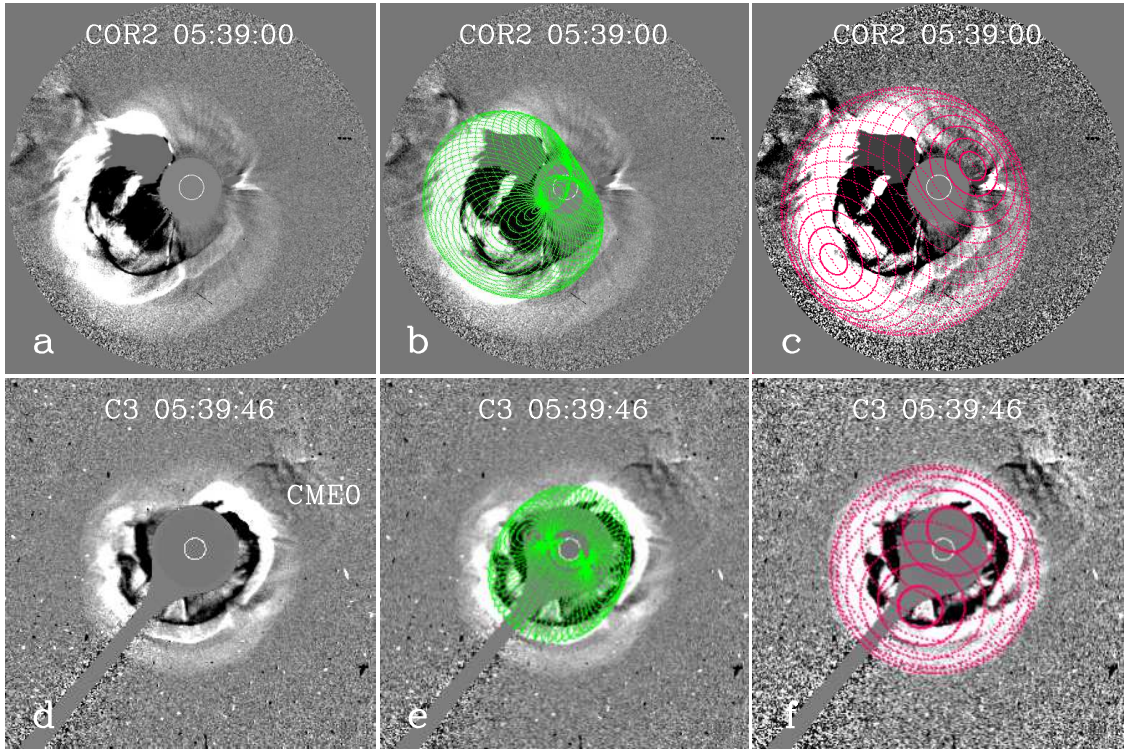} 
\caption{Multi-point imaging and modeling of the complex CME and shock. Top row: running-difference coronagraph image from STEREO A (a) and corresponding modeling of the CME (b) and shock (c). Bottom row: running-difference coronagraph image (d) and modeling (e, f) from SOHO near the same time. A sharper contrast is used in the right panels (c, f) to enhance the visibility of the shock.}
\end{figure}

\clearpage

\begin{figure}
\epsscale{0.6} \plotone{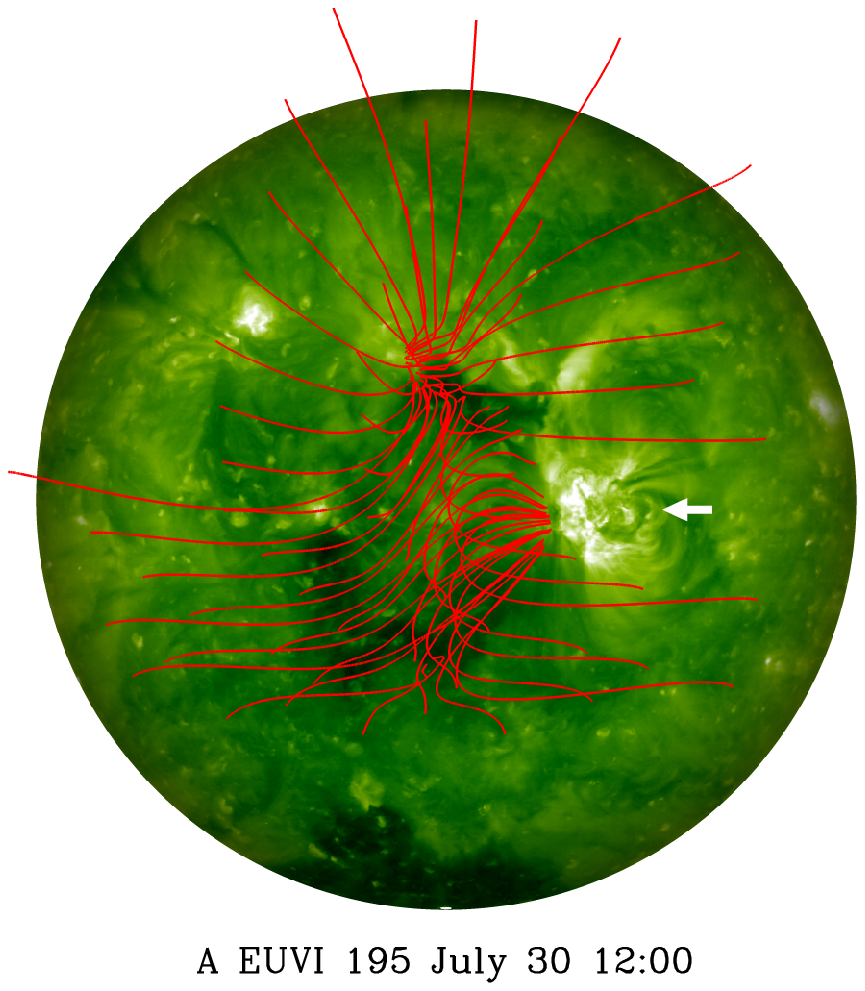} 
\caption{EUVI image at 195 \AA\ from STEREO A on 2017 July 30 together with open, directed inward magnetic field lines. The white arrow indicates the active region. Coronal holes with open field lines are seen on the immediate east of the active region.}

\end{figure}

\clearpage

\begin{figure}
\epsscale{0.65} \plotone{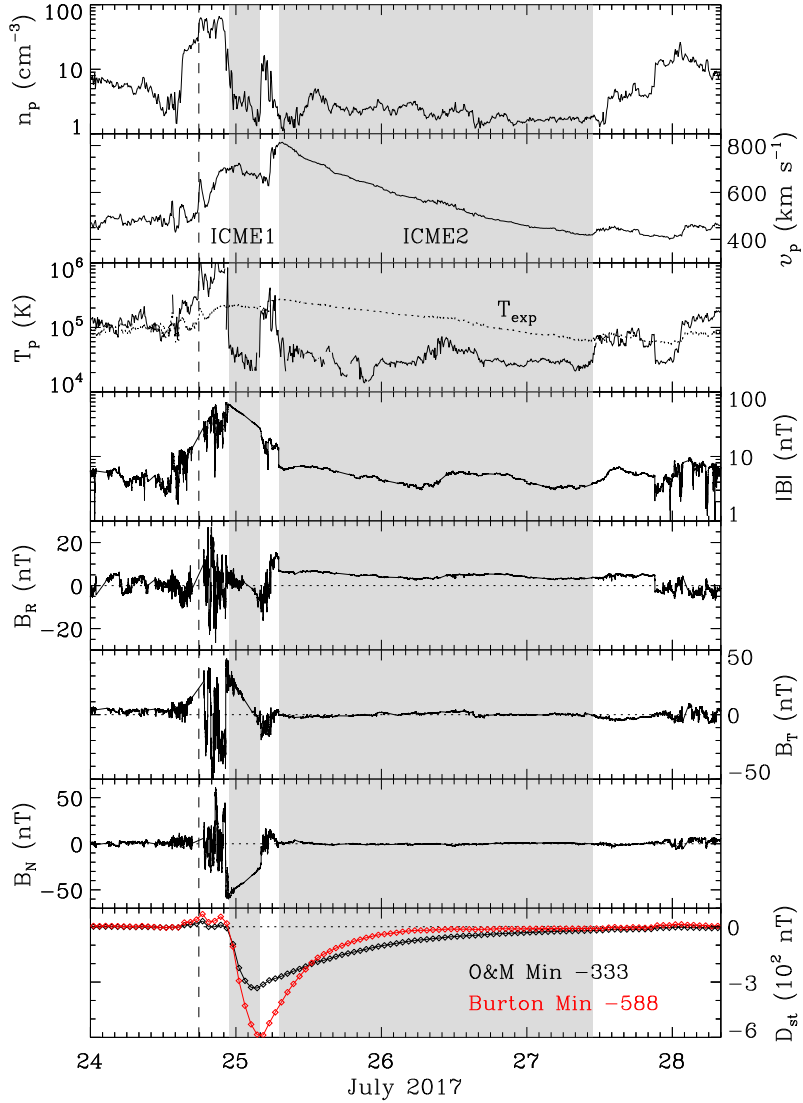} 
\caption{Solar wind measurements at STEREO A and calculated $D_{\rm st}$ index. From top to bottom, the panels show the proton density, bulk speed, proton temperature, magnetic field strength and components, and $D_{\rm st}$ index, respectively. The shaded regions indicate two ICME intervals, and the vertical dashed line marks shock arrival. The dotted curve in the third panel denotes the expected proton temperature derived from the observed speed \citep{lopez87}. The $D_{\rm st}$ values are estimated using the formulas of \citet[][red]{burton75} and \citet[][black]{om00}.}
\end{figure}

\clearpage

\begin{figure}
\epsscale{0.6} \plotone{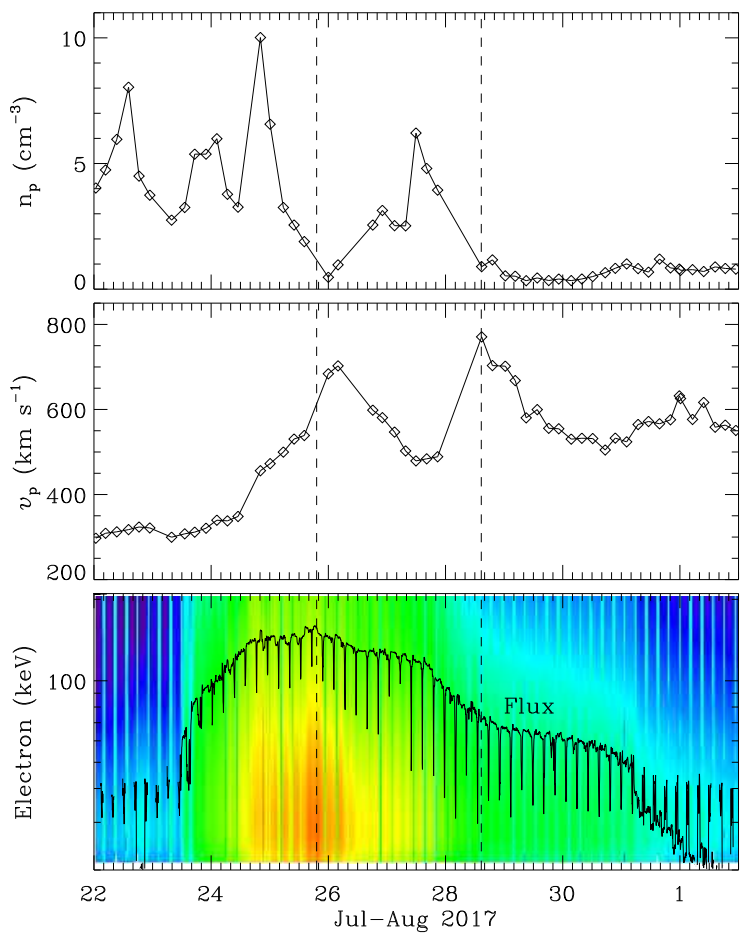} 
\caption{Solar wind measurements at Mars. From top to bottom, the panels show the proton density, bulk speed, and electron differential flux (descending from red to blue) at 20-210 keV. The black curve in the bottom panel represents a normalized flux averaged over the electrons. The sparse plasma parameters are averages of solar wind measurements over each 4.5-hour orbital period of MAVEN around Mars \citep{halekas17}. The 4.5-hour periodicity can also be seen in the electron data. The first vertical dashed line indicates a possible shock arrival determined from the peak of the electron flux, and the second one is likely the leading edge of ICME2.}
\end{figure}

\clearpage

\begin{figure}
\epsscale{0.95} \plotone{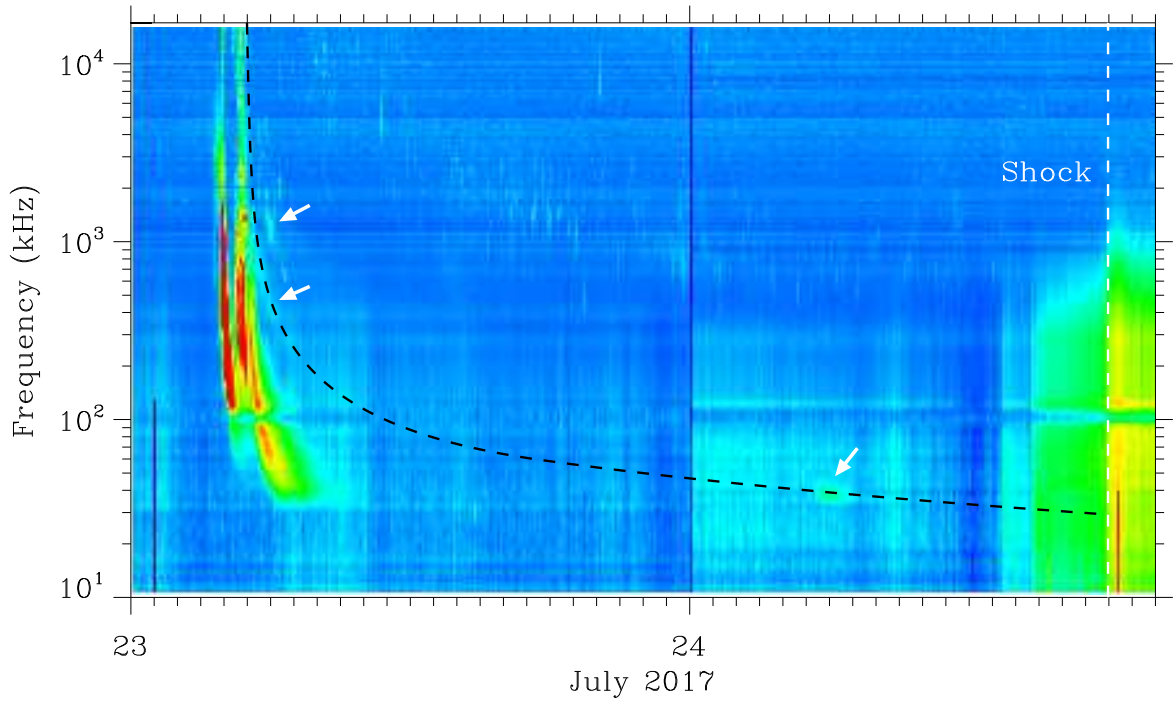} 
\caption{Dynamic spectrum from STEREO A. The white arrows indicate the type II bands. The black dashed curve, which is determined with a simple analytical model, simulates the frequency drift of the type II burst. The vertical dashed line denotes the shock arrival time from in situ measurements at STEREO A.}
\end{figure}

\end{document}